\documentclass[prb, reprint, twocolumn, showpacs, superscriptaddress]{revtex4-1}
\usepackage{graphicx} 
\usepackage{dcolumn}  
\usepackage{bm,bbm}       
\usepackage{amsfonts,amsmath,amssymb}
\usepackage{dsfont}
\usepackage{color}

\usepackage{soul,xcolor}
\usepackage{ulem}
\usepackage{color}

\usepackage{easyReview}
\begin{document}
\preprint{}
\title{High-Harmonic Generation from Engineered Graphene for Polarization Tailoring}

\author{Navdeep Rana}
\affiliation{%
	Department of Physics, Indian Institute of Technology Bombay,
	Powai, Mumbai 400076  India}

\author{M S Mrudul}
\affiliation{%
	Department of Physics and Astronomy, P.\,O.\ Box 516, Uppsala University, S-75120 Uppsala, Sweden.}

\author{Gopal Dixit}
\email{gdixit@phy.iitb.ac.in}
\affiliation{%
	Department of Physics, Indian Institute of Technology Bombay,
	Powai, Mumbai 400076  India}

\affiliation{%
	Center for Computational Sciences, University of Tsukuba, Tsukuba 305-8577, Japan.}
\affiliation{%
	Max-Born Institute, Max-Born Stra{\ss}e 2A, 12489 Berlin, Germany.}

\date{\today}

\begin{abstract}
Strain engineering is a versatile method to boost the carrier mobility of two-dimensional materials-based electronics and optoelectronic devices. In addition, strain is ubiquitous during device fabrication via material deposition on a substrate with a different lattice structure. 
Here, we show that the polarization properties of the harmonics in graphene under uniaxial strain are strongly yet differently affected in the lower and higher orders. 
The polarization plane of the lower-order emitted harmonics is rotated -- a manifestation of Faraday rotation due to the broken symmetry planes. 
In contrast, we observe elliptically-polarized higher-order harmonics due to the intricate interplay of the 
interband and intraband electron dynamics. 
The implications of these findings are twofold: First, we show how the rotation of the polarization plane of the lower-order harmonics can be used as a  probe to characterize the strain's nature, strength, and angle. 
Second, we demonstrate how strain engineering can be used to alter the polarization properties of higher-order harmonics, relevant for applications in ultrafast chiral-sensitive studies. Our research opens a promising avenue for strain-tailored polarization properties of higher-order harmonics in engineered solids.
\end{abstract}

\maketitle 
\section{Introduction}
Over the past decade, strong-field-driven high-harmonic spectroscopy has transpired as a robust method to interrogate numerous properties of solids on ultrafast timescale~\cite{ghimire2011observation,ghimire2019,goulielmakis2022high, luu2015extreme,   hohenleutner2015real, langer2018lightwave,pattanayak2019direct, imai2020high, borsch2020super, pattanayak2022role,  guan2022theoretical, vampa2015all, shao2019strain, kong2022strain, guan2019cooperative, tamaya2023shear, schmid2021tunable, qian2022role, bharti2022high, yan2023ultrafast, rana2022probing, kobayashi2021polarization, zuo2021neighboring, silva2018high, yoshikawa2017high, liu2017high, heide2022probing}.
In this respect, high-harmonic generation from the engineered solid targets is advantageous in two ways:
first, by actively controlling the generation process, the properties of the emitted harmonics can be tuned, and second, as the characterization method to investigate the properties of the engineered  targets~\cite{mcdonald2017enhancing, mrudul2020high,  liu2018enhanced, shcherbakov2021generation, pattanayak2020influence}. 
The present work demonstrates that the engineered two-dimensional (2D) materials not only increase the cutoff of the harmonics and enhance the yield but also offer the capability to tailor the harmonics' polarization. 
Necessary control over the polarization of the higher-order harmonics is quintessential to investigating a diverse range of chiral-sensitive light-matter phenomena~\cite{ferre2015table, kfir2015generation, boeglin2010distinguishing, giri2020time, radu2011transient,cireasa2015probing}. 
In addition, we will show that the analysis of the emitted harmonics allows us to diagnose the role of modified symmetries in the engineered 2D materials.  
	
Monolayer graphene's discovery has sparked enormous  interest in synthesizing a broad family of other 2D materials~\cite{novoselov2004electric, manzeli20172d}. 
Owing to the exceptional transport and optoelectronic properties,  2D materials have garnered tremendous attention in recent years as potential candidates for next-generation photonics and nanoelectronic devices~\cite{bonaccorso2010graphene, khan2022optical, rana2023all}. 
There have been continuous efforts to improve device performance  by tailoring the properties of the 
2D materials in a controlled manner~\cite{chaves2020bandgap, peng2020strain}. 
Strain engineering is a highly effective technique to control material's  properties as it has three control knobs: nature (tensile or compressive), direction (along zigzag, armchair and arbitrary), and strength of the strain~\cite{dai2019strain, naumis2017electronic, roldan2015strain}. 
Moreover, the presence of strain becomes unavoidable  when a 2D material is grown on a substrate with different structure and lattice parameters, e.g. during  the fabrication of devices like field-effect transistors~\cite{roy2014field, schwierz2015two, couto2014random}. 
Over the years, strained graphene has become a playground to improve carrier mobility in graphene-based devices as well as to explore new physics~\cite{bissett2014strain, amorim2016novel, galiotis2015graphene, guinea2010energy, si2016strain, frank2011development}. 
Thus, the characterization of  strain is of paramount importance for fundamental and technological applications.   
In recent years, the impact of strain on  solids  has been studied in the context of high-harmonic generation~\cite{shao2019strain,guan2019cooperative,kong2022strain,tamaya2023shear,qin2018strain}. 
These investigations have demonstrated that strain leads to an  enhancement or a quenching of the harmonic yield. 
	
Graphene under strain is chosen to illustrate  the polarization tunability of the emitted harmonics.
Moreover, analysis of the emitted harmonics allows us to characterize strain strength {\it as small as}  0.001, which corresponds to 0.05 picometer change in bond distance  between two atoms in graphene. 
Owing to high flexibility under external stimuli, various methods have been used to introduce strain in monolayer graphene, such as the application of mechanical pressure via STM tips~\cite{klimov2012electromechanical}, 
via  gas inflation~\cite{bunch2008impermeable} to name but a few~\cite{lee2008measurement, guinea2012strain, zhang2022strain}. 
Moreover, lattice and/or thermal expansion  mismatch during  growth on a substrate can also introduce strain~\cite{tomori2011introducing, bao2009controlled}. 
Polarized Raman spectroscopy is routinely employed for quantifying strain  in graphene by analyzing  the shift in vibrational  bands~\cite{huang2009phonon, ferrari2013raman, wu2018raman}. 
However, sensing the strain strength below 0.1$\%$  is challenging as the shift in the 2D peak position is much smaller than its peak width~\cite{del2015strain, tomori2020improved}. 
Such a limitation is not  the case for high-harmonic spectroscopy.

In the following, we will show how the tensile and compressive uniaxial  strains affect high-harmonic generation in graphene differently.  
Polarization of the lower-order harmonics is observed to be rotated with respect to the driving laser.
The rotation of the polarization plane of the lower-order harmonics  is analogous to the optical Hall effect in strained graphene -- Faraday rotation~\cite{guinea2010energy}. 
Information about the strength,  direction, and nature of the uniaxial strain  can be extracted by analyzing the  polarization dependence of the harmonics.
On the other hand, the higher-order harmonics are elliptical in nature, and there ellipticity can be tuned by  the nature, strength, and direction of the strain.

\section{Theoretical Methodology}
The strain tensor corresponding to a  uniaxial strain on graphene is described as~\cite{pereira2009tight, pereira2009strain} 
\begin{equation}
\mathcal{\bm{\varepsilon}} = \varepsilon
\begin{bmatrix}
\cos^{2}\theta_{s} - \sigma \sin^{2}\theta_{s} && (1+\sigma)\cos \theta_{s} \sin \theta_{s} \\
(1+\sigma)\cos \theta_{s} \sin \theta_{s} && \sin^{2}\theta_{s} - \sigma \cos^{2}\theta_{s}
\end{bmatrix}.
\label{eq:strain_tensor}
\end{equation} 
Here, $\varepsilon$ quantifies the strain's strength, and  $\theta_{s}$ is the  direction  along which the strain is applied and measured with respect to the zigzag direction ($\mathsf{X}$-axis).
$\sigma$ = 0.165  is the Poisson's ratio~\cite{blakslee1970elastic}. 
Under the uniaxial strain of strength $\varepsilon$ along $\theta_{s}$, any undeformed vector $\mathbf{u}$ in the real-space is transformed as 
\begin{equation}
\tilde{\bm{u}} =  (\mathbbm{1} + \mathcal{\bm{\varepsilon}}) \cdot \bm{u},
\label{eq:transform}
\end{equation}
where $\mathbbm{1}$ is the 2$\times$2 identity matrix.

The tight-binding approach is used to describe the monolayer graphene as~\cite{reich2002tight}
\begin{equation}
\mathcal{H}_\mathbf{k}  = - \sum_{<mn>}  \gamma_{mn}  ~e^{i\textbf{k}\cdot \bm{\delta}_{mn}} \hat{a}_{m\textbf{k}}^{\dagger}\hat{b}_{n\textbf{k}}+\text{H.c.}
\label{eq:tb}
\end{equation}
Here, $\hat{a}_{m\textbf{k}}^{\dagger}~(\hat{a}_{m\textbf{k}})$  represents the electronic creation (annihilation) operator  at the A sublattice, while  $\hat{b}_{n\textbf{k}}^{\dagger}~(\hat{b}_{n\textbf{k}})$  denotes the electronic creation (annihilation) operator at the B sublattice. The summation is performed  over the nearest-neighboring  atoms. 
An exponential  function of inter-atomic distance describes the hopping energy as $\gamma_{mn} = \gamma_{0} e^{-(\delta_{mn}-a)/b}$ with   
$\bm{\delta}_{mn}$ as the separation vector between nearest-neighbor atoms, $\gamma_{0} = $ 2.7 eV,   $a$ =  1.42 \AA~ as the inter-atomic distance, and $b$ = 0.32$a$ as the decay length~\cite{pereira2009tight,moon2013optical}. 
When the uniaxial strain is applied in graphene,  atomic positions, and the lattice vectors are modified according to Eq.~(\ref{eq:transform}). 
Moreover,  the Hamiltonian in Eq.~(\ref{eq:tb})  changes accordingly. 
In addition, Fermi level remains unchanged after the application of the strain 
on graphene.  
Density-matrix-based semiconductor Bloch equations are solved numerically to simulate laser-driven  electron dynamics in strained graphene~\cite{rana2022generation,rana2022high}. 
The Fourier transform ($\mathcal{FT}$) of the time-derivative of the charge current is used to obtain the harmonic spectrum   as $|\mathcal{FT}(\partial \textbf{J}(t)/\partial t)|^2$, where $\textbf{J}(t)$ stands for the charge current generated by the driving laser~\cite{mrudul2021high}. 
A constant phenomenological dephasing time of 10 fs is used, and our finding remains same for varying dephasing time from 2 to 20 fs (see supplemental Figs. S1-S2).
A 85 fs  long  linearly polarized pulse with 3.2 $\mu$m wavelength and 10$^{11}$ W/cm$^2$ peak intensity  is employed. 
Our findings remain qualitatively same for laser intensity ranging from 0.8 to 1.4$\times$10$^{11}$ W/cm$^2$,
and wavelength variation from 2.0 to 4.8 $\mu$m.  
The driving laser is polarized  along the $\mathsf{X}$ direction throughout this paper unless stated otherwise.
Here,  armchair and zigzag directions are aligned with the $\mathsf{Y}$ and $\mathsf{X}$ axes, respectively. 

\section{Results and Discussion}
\begin{figure} 
\includegraphics[width= \linewidth]{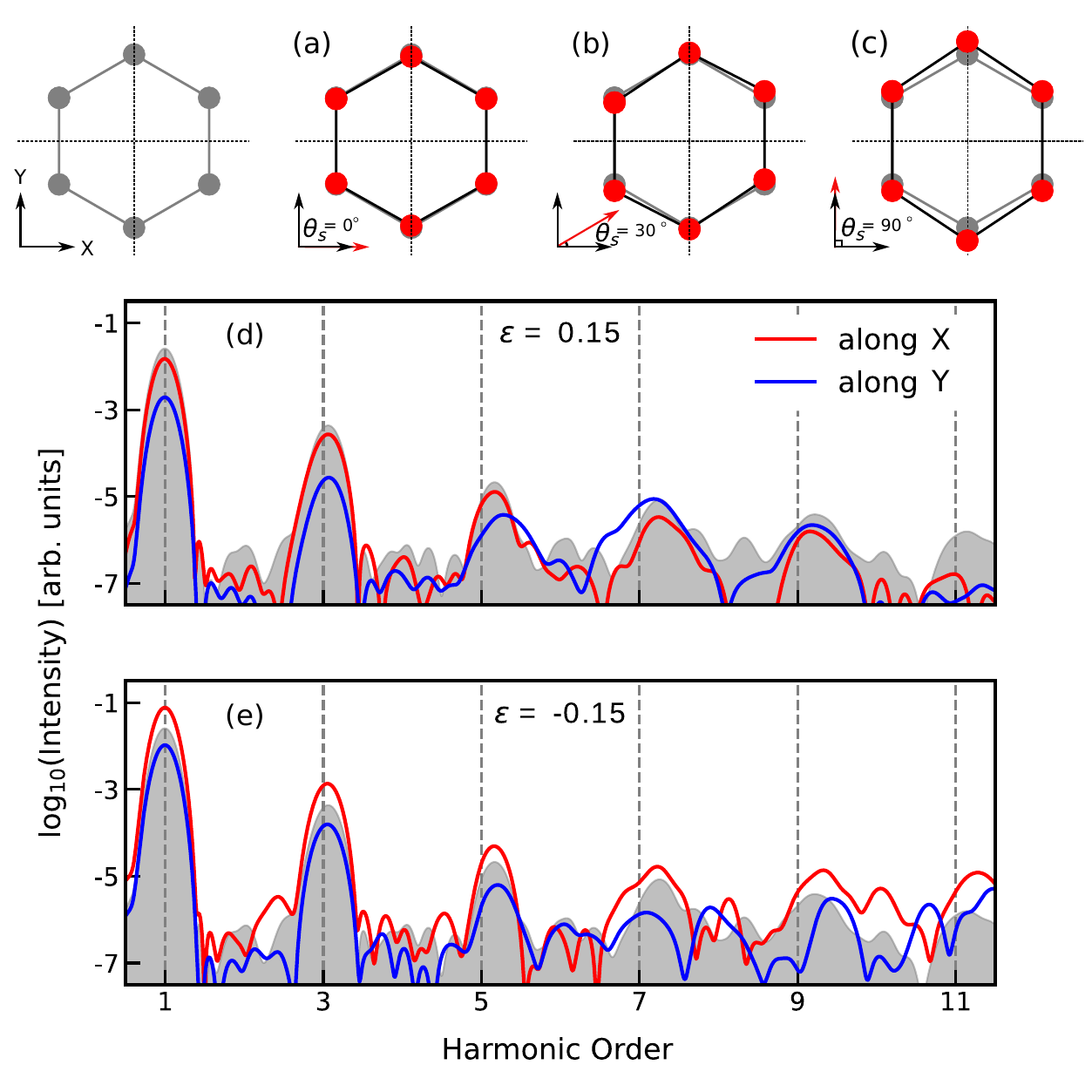}
\caption{Real-space structure of the strained graphene when the uniaxial tensile strain is applied along 
(a) zigzag, (b) 30$^{\circ}$, and (c) armchair directions. 
The structure of the graphene without strain is shown in the grey as a reference structure.  
High-harmonic spectra of the strained graphene for strain strengths (d) $\varepsilon = 0.15$ and (e) $\varepsilon = -0.15$ along $\theta_{s} = 30^{\circ}$. 
The red and blue colors stand for the polarization of emitted harmonics along the $\mathsf{X}$ (zigzag) and $\mathsf{Y}$ (armchair) directions, respectively. 
The driving laser pulse is polarized along the $\mathsf{X}$ direction.
The grey-shaded region represents the high-harmonic spectra of unstrained graphene,  serving as the reference. The emitted harmonics are polarized along the $\mathsf{X}$ direction for unstrained graphene.}\label{spectra_disper}
\end{figure}

Let us discuss how microscopic modifications caused by the uniaxial strain impact high-harmonic generation. 
Figure~\ref{spectra_disper} present the high-harmonic spectra of graphene under the  tensile and  compressive  strains along $\theta_{s} = 30^{\circ}$ direction, and the respective real-space lattice structures for different $\theta_{s}$ values.  
The lattice structure of the unstrained graphene ($\varepsilon = 0.0$), and the corresponding high-harmonic spectrum are shown as references in the grey color.  
The uniaxial strain leads to  noticeable changes in the  spectra. 
Moreover, the tensile and compressive strains affect the harmonic generation distinctively as visible from Figs.~\ref{spectra_disper}(d) and ~\ref{spectra_disper}(e), respectively. 	
In the case of tensile strain, the lattice stretches along the strain direction and compresses in the perpendicular direction as governed by Poisson's ratio.
This situation is opposite when the strain is compressive in nature [see Eqs.~(\ref{eq:strain_tensor} - \ref{eq:transform})]. 
The different behavior of the tensile and compressive strains are evident from the asymmetric nature of the Hamiltonian with respect to $\varepsilon$, described in Eq.~(\ref{eq:tb}). Thus, the emitted harmonics are sensitive to the nature of the applied strain.

The absence of even-order harmonics indicates that the uniaxial strain preserves the inversion symmetry in strained graphene. 
Interestingly, the generated harmonics are polarized parallel and perpendicular to the driving pulse's polarization. This follows the broken reflection  symmetries  along  $\mathsf{XZ}$  and $\mathsf{YZ}$ planes  when graphene is stretched uniaxially along $\theta_{s} = 30^{\circ}$ with respect to the zigzag direction as
evident from Fig.~\ref{spectra_disper}(b).
The perpendicular component of higher-order harmonics is absent when strain is applied along zigzag [Fig.~\ref{spectra_disper}(a)] or armchair [Fig.~\ref{spectra_disper}(c)] directions, as they preserve the XZ and YZ reflection planes, analogous to unstrained graphene.
At this juncture,  it is interesting to explore how the generation of the parallel and perpendicular components affects the polarization of the emitted harmonics. 
To address this crucial question, let us first focus on the low-order harmonics.

\begin{figure}
\includegraphics[width=\linewidth]{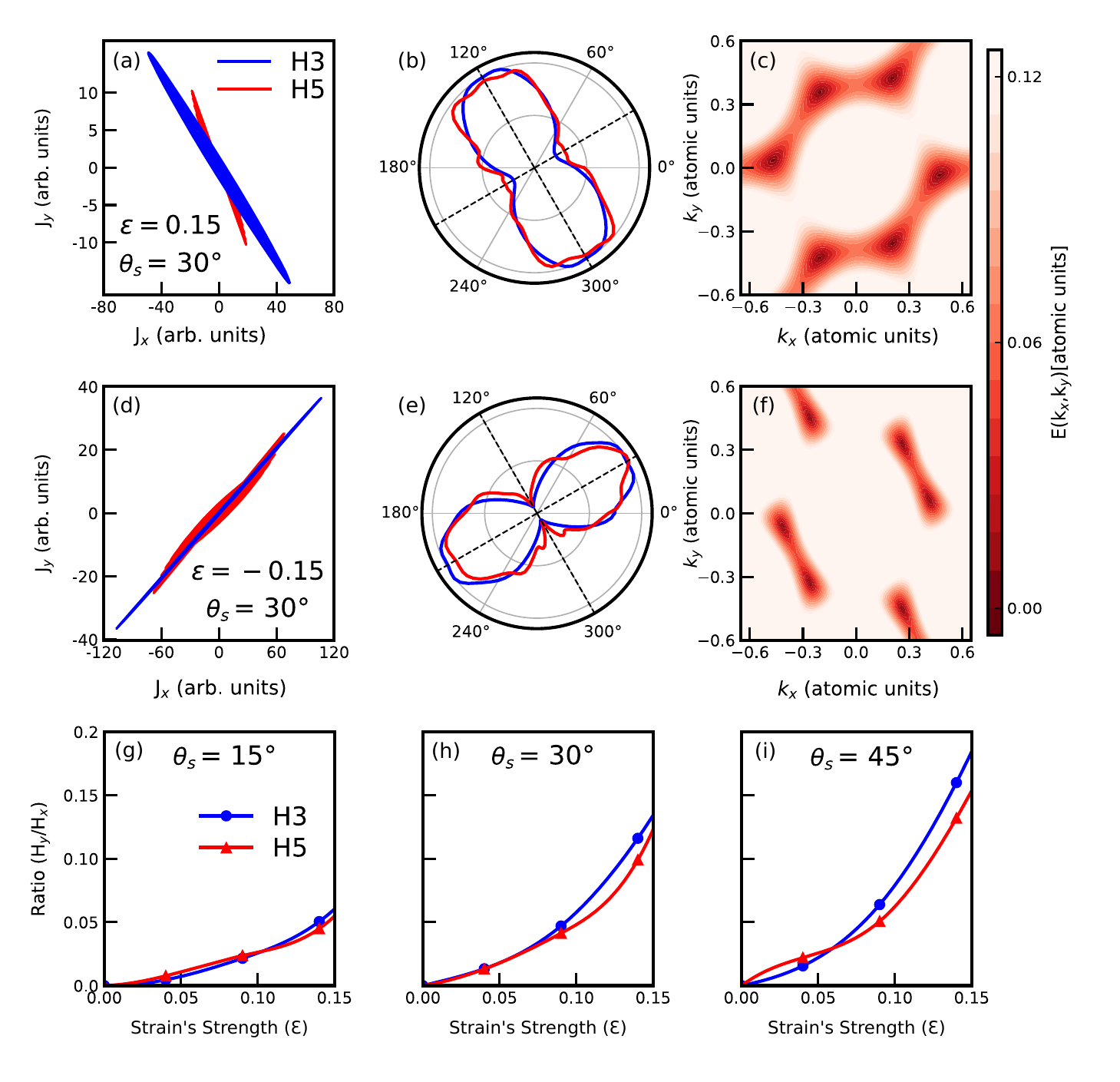}
\caption{Projection of the time profile of the third (H3) and fifth (H5) harmonics corresponding to 
graphene under uniaxial strain with  (a) $\varepsilon$ = 0.15 and (d) $\varepsilon$ = - 0.15 along 
$\theta_{s}$ = 30$^\circ$.  
Variations in the normalized harmonic yields of H3 and H5 for 
(b) $\varepsilon$ = 0.15 and (e) $\varepsilon$ = - 0.15 along  $\theta_{s}$ = 30$^\circ$.
Energy contour of the conduction band of graphene under the uniaxial strain of 
strength (c) $\varepsilon$ = 0.15 and (f) $\varepsilon$ = - 0.15  along  $\theta_{s}$ = 30$^{\circ}$. 
The ratio of the perpendicular and parallel components of the total harmonic yield for graphene under different values of $\varepsilon$ along (g) $\theta_{s}$ = 15$^\circ$, (h) 30$^\circ$, and (i) 45$^\circ$.}
\label{yield}
\end{figure}

Figures~\ref{yield}(a) and \ref{yield}(d) present projection of the time profile of the low-order harmonics corresponding to the spectra shown in 
Figs.~\ref{spectra_disper}(d) and ~\ref{spectra_disper}(e), respectively. 
In-phase relation between the parallel and perpendicular components of the total current, corresponding to third (H3) and fifth (H5) harmonics,  causes a rotation of the polarization plane in strained graphene. 
The rotation of the polarization plane in strained graphene can be attributed to broken reflection symmetries along the $\mathsf{XZ}$  and $\mathsf{YZ}$ planes, which leads to the generation of  the current perpendicular to the driving laser field's polarization.  
This is analogous to the optical Hall effect~\cite{nguyen2017optical}.  
The rotation is  in the opposite direction for the tensile and compressive strains as evident from Figs.~\ref{yield}(a) and \ref{yield}(d), respectively. 
Thus, the rotation is sensitive to the nature of the strain.  
	
Reliable information about  the strain's direction can not be extracted exclusively 
from  Figs.~\ref{yield}(a) and \ref{yield}(d) for the following reasons: 
Unstrained graphene exhibits 30$^\circ$ and 90$^\circ$ as symmetrically equivalent directions 
from its six-fold rotational symmetry. This changes for strained graphene.  
There is no change  in the polarization plane when the strain is along 
90$^\circ$ due to the presence of reflection plane symmetry  [see Fig.~\ref{spectra_disper}(c)], 
which contrasts the situation when strain is along 30$^\circ$ for the laser polarized along $\mathsf{X}$ direction 
[see Fig.~\ref{spectra_disper}(b)].  
Thus,  the extent  of the polarization-rotation depends simultaneously on the strain's angle and 
the polarization of the driving pulse.  
Therefore, it is imperative to analyze the laser polarization direction dependence on the polarization  properties of the emitted harmonics.
	
Figures~\ref{yield}(b) and \ref{yield}(e) present 
polarization dependence of H3 and H5 for   $\varepsilon$ = 0.15 and  $\varepsilon$ = - 0.15 along  
$\theta_{s} = $  30$^{\circ}$, respectively.   
Both H3 and H5 are highly anisotropic in nature  and the maximum yield is sensitive to the direction of the applied  
strain.  On the other hand,  the polarization dependence of H3 corresponding to  
the unstrained graphene is isotropic, whereas H5  exhibits six-fold symmetry as reported earlier~\cite{mrudul2021high}.  
It is striking to observe that  the direction of the maximum yield is perpendicular to the direction of the tensile strain. 
This observation changes when the strain's nature  changes from tensile to compressive. 
The direction of maxima is along the direction of applied strain as evident from Fig.~\ref{yield}(e). 
Thus, it can be concluded that the maxima of the harmonic yield align along the direction in which the lattice structure is compressed [see Figs.~\ref{spectra_disper}(a)-\ref{spectra_disper}(c)]. The principal axis of the unstrained graphene exhibits six-fold symmetry, which reduces to two-fold 
for graphene under uniaxial strain [see Figs.~\ref{spectra_disper}(a)-\ref{spectra_disper}(c)], resulting in the anisotropic behavior. 
The modifications in the conduction band corresponding to $\varepsilon$ = 0.15 and  $\varepsilon$ = - 0.15 are, respectively, shown in Figs.~\ref{yield}(c) and \ref{yield}(f). The distortions in the band structure drive electrons to different parts of the Brillouin zone depending on the angle and nature of the strain, recording the microscopic structural alterations in the polarization map.
		
So far, we have discussed results for $\varepsilon  = \pm 0.15$. 
Let us investigate how the polarization rotation changes with $\varepsilon$. 
For this purpose, we analyze the ratio of the perpendicular and parallel harmonic yield for different  $\varepsilon$. 
Figures~\ref{yield}(g)-\ref{yield}(i) present the ratio for H3 and H5 as a function of  $\varepsilon$ and  $\theta_{s}$. 
It is interesting to observe that the ratio increases monotonically with $\varepsilon$. 
Moreover, the ratio is also sensitive to $\theta_{s}$. 
It indicates that the rotation of the polarization plane has a nonlinear increase with $\varepsilon$ and $\theta_{s}$. 
Thus, one could qualitatively quantify $\theta_{s}$  by analyzing the ratio of the harmonic components. 

\begin{figure}
\includegraphics[width=\linewidth]{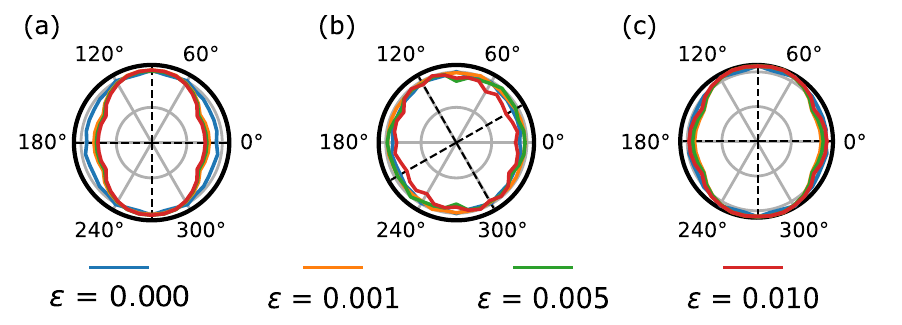}
\caption{Polarization-dependence on the normalized  yield of H3 in strained graphene exposed to different strengths of the tensile uniaxial strain
along (a) $\theta_{s}$ = 0$^{\circ}$,  (b) 30$^{\circ}$, and (c) 90$^{\circ}$. }
\label{yield_angle}
\end{figure}
	
We further comment on the sensitivity of the polarization rotation with respect to $\varepsilon$.
It is interesting to observe that the strength as small as  $\varepsilon = 0.01$ results  in a significant anisotropic dependence. 
Moreover, a measurable anisotropy is present even for $\varepsilon = 0.001$, which corresponds to a 0.03 $\%$ change in the lattice parameter [see Fig.~\ref{yield_angle}]. 
Thus, the extent of the anisotropy can be used to quantify  the strain's strength. 
Moreover, the direction and nature of the strain can be extracted from the maxima of the anisotropic yield. 
It is interesting to note that the maximum harmonic yield is along the stretching direction when the uniaxial tensile strain is along 90$^{\circ}$ [Fig.~\ref{yield_angle}(c)]. 
However,  we observe the harmonic yield is peaked along the direction of compression when $\epsilon$ is greater than 0.02 (see supplemental Figs. S3-S4), which is analogous to Fig.~\ref{yield}(b).
Therefore, high-harmonic spectroscopy can be seen as a robust method to characterize  the nature, direction, and strength of the uniaxial strain in graphene. 
	
\begin{figure}
\includegraphics[width= 0.85\linewidth]{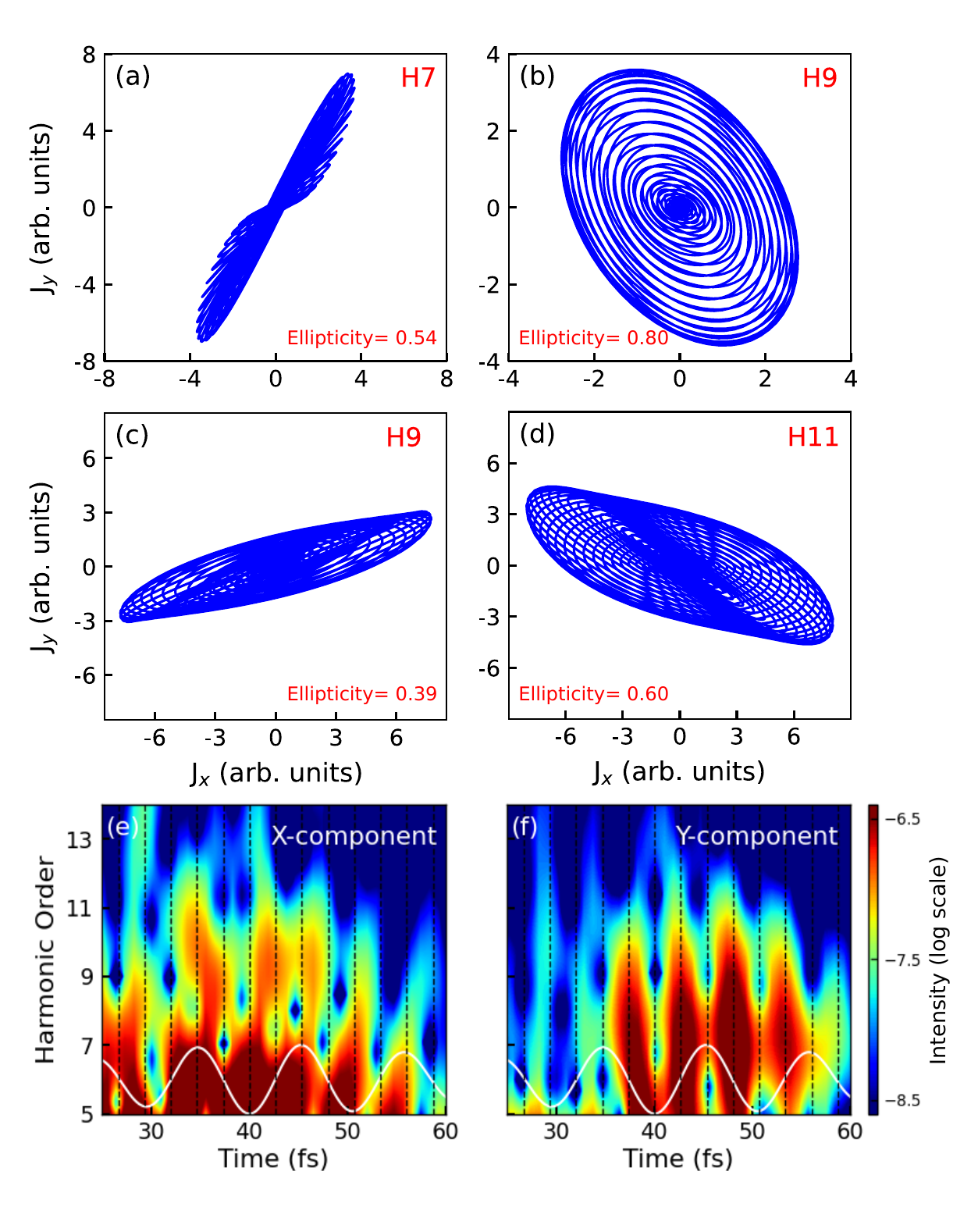}
\caption{Projection of time profile of the (a) seventh (H7) and (b) ninth (H9) harmonics of  the 
strained graphene with $\varepsilon $ = 0.15  for $\theta_{s}$ =30$^{\circ}$. 
(c) and (d) same as (a) and (b), respectively, for ninth (H9) and eleventh (H11) harmonics with $\varepsilon $ = - 0.15.
Time-frequency map of the (e) $\mathsf{X}$- and (f) $\mathsf{Y}$-components of the interband harmonics
of the strained graphene with $\varepsilon $ = 0.15  along $\theta_{s}$ =30$^{\circ}$. 
The electric field of the driving laser pulse is shown in white and  
the black dashed lines correspond to minima, maxima, and zeros of the electric field.}
\label{Polarizion}
\end{figure}

Let us shift our attention toward the properties of the higher-order harmonics. 
Figures~\ref{Polarizion}(a) and \ref{Polarizion}(b) display 
the temporal profiles of the seventh (H7) and ninth (H9) harmonics of strained graphene 
with $\varepsilon$ = 0.15 along $\theta_{s}$ = 30$^{\circ}$, respectively. 
In contrast to  the linear polarization of H3 and H5, H7 and H9 are  elliptical as evident from the figure. 
Similarly,  ninth (H9) and eleventh (H11) harmonics of the strained graphene, with
$\varepsilon$ = -0.15 along $\theta_{s}$ = 30$^{\circ}$, are elliptical as reflected from 
Figs.~\ref{Polarizion}(c) and \ref{Polarizion}(d), respectively. 
It is fascinating to note that the polarization properties of the harmonics are contrasting in the lower and higher 
orders in strained graphene. 
In the lower orders, we only observed a rotation of the polarization plane of the emitted harmonics. 
On the other hand, we observe a out of phase relation
in the parallel and perpendicular components, resulting in the elliptically polarized higher-order harmonics.
		
To have a  better understanding of the unusual  polarization of higher-order harmonics, let us analyze the time-frequency map of the harmonics of the strained graphene with $\varepsilon $ = 0.15  along $\theta_{s}$ = 30$^{\circ}$. 
Analysis of the time-frequency map of the $\mathsf{X}$-component of the interband harmonics indicates that the harmonic bursts are around the crest and trough of the electric field as evident from Fig.~\ref{Polarizion}(e). 
On the other hand,  $\mathsf{Y}$-component demonstrates that the  harmonic bursts occur around the zeros of the electric field as reflected from Fig.~\ref{Polarizion}(f). 
Thus, an out of the  phase relation between the $\mathsf{X}$- and $\mathsf{Y}$-components is observed, indicating different underlying mechanisms of electron dynamics.

Let us further delve into the physical process responsible for the anisotropic and unusual polarization dependence. 
It is known that the polarization properties of the emitted harmonics strongly depend on the nature of the underlying electron dynamics~\cite{tancogne2017ellipticity}. 
An intricate interplay of the interband and intraband contributions in higher-order harmonic generation in graphene was shown previously~\cite{mrudul2021high}. 
In the following part, we explain the unusual polarization properties of higher-order harmonics in strained graphene by analyzing the interband and intraband components of the harmonics.

The charge current, $\textbf{J}(t)$, can be estimated with the density-matrix element ($\rho_{mn}^\textbf{k}$) and momentum-matrix element ($\textbf{p}_{mn}^\textbf{k}$) as
\begin{equation}
\begin{split}
\textbf{J}(t) &\propto \sum_{m \neq n,\textbf{k}}\rho_{mn}^\textbf{k}(t)\textbf{p}_{nm}^{\textbf{k}+\textbf{A}(t)} + \sum_{m,\textbf{k}}\rho_{mm}^\textbf{k}(t)\textbf{p}_{mm}^{\textbf{k}+\textbf{A}(t)} \\	
&= \textbf{J}_{inter}(t) + \textbf{J}_{intra}(t) .
\end{split}
\end{equation}
Here, $\textbf{A}(t)$ is the vector potential of the laser field.  $\textbf{J}_{inter}(t)$ and  $\textbf{J}_{intra}(t)$ are the 
interband and intraband currents, respectively. The ultrafast electron dynamics can be through either of these 
contributions. 

\begin{figure}
\includegraphics[width=\linewidth]{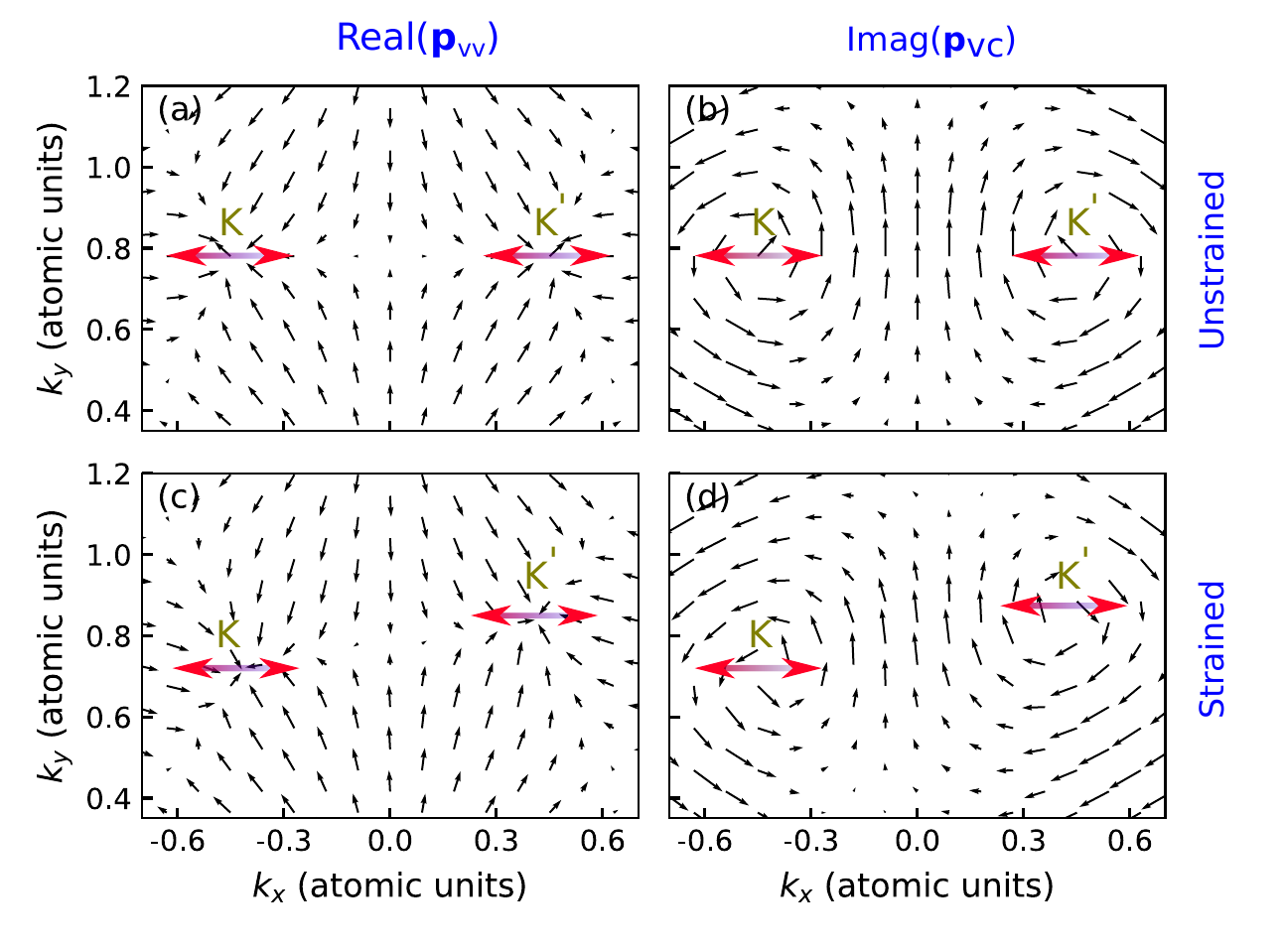}
\caption{Diagonal momentum matrix element of the valence band (band-velocity, $\textbf{p}_{vv}$) of (a) unstrained graphene and (c) strained graphene with $\varepsilon$ = 0.15 and $\theta_{s}$ = 30$^\circ$. 
The off-diagonal momentum matrix element between valence and conduction bands ($\textbf{p}_{vc}$) 
of (b) unstrained and (d) strained graphene. $\mathsf{K}$ and $\mathsf{K}^{\prime}$ are high-symmetry points in the Brillouin  zone.}
\label{fig:compare}
\end{figure}

The diagonal and off-diagonal components of momentum-matrix element are shown in Fig.~\ref{fig:compare}. 
The diagonal elements of the momentum-matrix element are the band-velocities, and the off-diagonal elements are proportional to the dipole coupling. 
Note that, according to the semiclassical description of the electron dynamics, the intraband current is proportional to the group velocity and Berry curvature. 
Strained graphene exhibits inversion and time-reversal symmetries, and therefore Berry curvature is zero. 
Another consequence of having these symmetries is pure imaginary off-diagonal momentum-matrix elements [Figs.~\ref{fig:compare}(b) and \ref{fig:compare}(d)].

It is evident that the off-diagonal momentum matrix elements close to distinct K-points have opposite chirality [see Fig.~\ref{fig:compare}(b)].
Thus, the interband currents generated close to different K-points are out of phase and interfere destructively. Also, the available joint density of states for interband transitions are less in the lower energy response~\cite{mrudul2021high}.
So, the lower-order harmonics are dominantly from the intraband contributions, which constructively interfere 
from $\mathsf{K}$ and $\mathsf{K}^{\prime}$ regions as evident from Fig. ~\ref{fig:compare}(a). 
The unstrained graphene favors the parallel component of the intraband current, apparent from the symmetries of the band velocity [see the reflection symmetry with respect to X-axis for Y-components of the vector field in Fig.~\ref{fig:compare}(a)].  
In strained graphene, the reflection planes are broken, and the vector field is distorted [see Fig.~\ref{fig:compare}(c)], resulting in the perpendicular component of the intraband current. 
As the parallel and perpendicular currents are generated from the same underlying mechanism,  they are in phase, resulting in the rotation of the plane of polarization of the lower-order harmonics [Fig.~\ref{yield}(a)]. 
Moreover, these distortions in the band velocities are responsible for the characteristic polarization-dependence 
observed in Fig.~\ref{yield}(b).
	
When electrons are excited away from K-points along the laser polarization direction, 
they result in a perpendicular contribution of interband current [Figs.~\ref{fig:compare}(b) and \ref{fig:compare}(d)]. 
However, the perpendicular component in unstrained graphene cancels among different K-points 
due to the symmetries of the vector field [Fig.~\ref{fig:compare}(b)]. 
Since the symmetry planes are absent in strained graphene, 
a net perpendicular interband current is generated from the distorted vector field [Fig.~\ref{fig:compare}(d)]. 
It is essential to consider that the parallel and perpendicular current contributions happen from different regions in the Brillouin zone [see the modification in the Y-component of the momentum matrix elements throughout the Brillouin zone in Fig.~\ref{fig:compare}(c) and \ref{fig:compare}(d)]. 
So, different non-linear processes contribute in parallel and perpendicular directions. The interference of different non-linear processes can result in different time-frequency maps in the orthogonal directions, generating a phase difference and resulting in elliptical harmonics.
In short, this nontrivial interplay of intraband and interband currents in the parallel and perpendicular directions is responsible for the elliptical higher-order harmonics as observed in Figs.~\ref{Polarizion}(a)-(d).

\section{Conclusion}
To summarize, we have unequivocally demonstrated how the analysis of the polarization-dependence of
the emitted harmonics can be used to quantify both the qualitative and quantitative nature of the strain in graphene. 
It is observed that the uniaxial strain in graphene affects the harmonic generation drastically.  
In addition, tensile and compressive uniaxial strains influence the generation process differently. 
The absence of the reflection symmetries in strained graphene leads to the generation of harmonics perpendicular to the incident laser pulse.
The $\mathsf{X}$- and $\mathsf{Y}$-components of the lower-order harmonics are in phase, which results
rotation of the polarization plane -- analogous to the optical Hall effect. 
The extent of the  rotation depends on the angle at which the strain is present, 
and exhibits a non-linear scaling with the strain's strength. 
Higher-order harmonics have an intricate  interplay of the interband and intraband contributions. 
The $\mathsf{X}$- and $\mathsf{Y}$-components of the higher-order harmonics are in  out of phase relation, 
which leads to the generation of elliptically or circularly polarized harmonics.
Our findings add a new dimension to high-harmonic spectroscopy by establishing sensitivity toward strain, which will have important technological fall-outs while designing and characterizing 2D materials-based engineered devices where strain is unavoidable~\cite{sivis2017tailored, korobenko2022situ, abbing2022extreme}.
Moreover, controlling the properties of engineered graphene allows us to tailor the polarization of emitted harmonics from linear to elliptical, which will be useful to interrogate numerous chiral-sensitive light-matter interactions on ultrafast timescale.

\section*{Acknowledgments}
We acknowledge fruitful discussions with Misha Ivanov (MBI, Berlin), David Reis (SLAC, USA), Akinobu Kanda (Tsukuba University), and Sumiran Pujari (IITB). G.D. acknowledges financial support from SERB India (Project No. MTR/2021/000138).  

\bibliography{solid_HHG}

\end{document}